\documentclass[twocolumn]{article}
\usepackage{parskip}
\usepackage{amssymb}
\usepackage{amsmath}
\usepackage{fullpage}
\usepackage{authblk}
\usepackage{blindtext}
\usepackage{siunitx}
\usepackage{float}
\usepackage{caption}
\usepackage[superscript,nomove]{cite}
\usepackage{hyperref}
\usepackage[normalem]{ulem}
\usepackage{graphicx}
\usepackage{cancel}

\makeatletter \renewcommand{\@citess}[1]{\textsuperscript{[#1]}} \makeatother

\makeatletter
\def\@firstoftwo@second#1#2{%
  \def\temp##1.##2\@nil{##2}%
   \temp#1\@nil}
\newcommand\sref[1]{%
   (A.\expandafter\@setref\csname r@#1\endcsname\@firstoftwo@second{#1})%
}

\makeatother

\title{Prediction of defect properties in concentrated solid solutions using a Langmuir-like model}
\author[1,2]{\small Jacob Jeffries \thanks{jwjeffr@g.clemson.edu}}
\author[4]{Fadi Abdeljawad}
\author[5]{Suveen Mathaudhu}
\author[6]{Emmanuelle Marquis}
\author[1,3]{Enrique Martinez \thanks{enrique@clemson.edu}}
\affil[1]{Department of Materials Science and Engineering, Clemson University, Clemson, SC 29634, USA}
\affil[2]{Theoretical Division, Los Alamos National Laboratory, Los Alamos, NM 87545, USA}
\affil[3]{Department of Mechanical Engineering, Clemson University, Clemson, SC 29634, USA}
\affil[4]{Department of Materials Science and Engineering, Lehigh University, Bethlehem, PA 18015, USA}
\affil[5]{Metallurgical and Materials Engineering Department 
Colorado School of Mines, Golden, CO 80401, USA}
\affil[6]{Material Science and Engineering 
University of Michigan, Ann Arbor, MI 48109 USA}
\date{\small \today}

\renewenvironment{abstract}
 {\quotation\small\noindent\rule{\linewidth}{.5pt}\par\smallskip
  {\centering\bfseries\abstractname\par}\medskip}
 {\par\noindent\rule{\linewidth}{.5pt}\endquotation}

\begin{document}

\twocolumn[
  \begin{@twocolumnfalse}
  \maketitle
    \begin{abstract}
        The alleged existence of sluggish diffusion in high entropy alloys has drawn controversy. In high entropy alloys, and in general in all solids, transport properties are controlled by point defect concentration, which must be known before performing atomistic simulations to compute transport coefficients. In this work, we present a general Langmuir-like model for defect concentration in an arbitrarily complex solid solution and apply this model to generate expressions for concentrations of vacancies and small interstitial atoms. We then calculate the vacancy concentration as a function of temperature in the equiatomic CoNiCrFeMn and FeAl alloys with modified embedded-atom-method potentials for various chemical orderings, showing there is no clear correlation between vacancy thermodynamics and chemical ordering in the CoNiCrFeMn alloy but clear systematic patterns for FeAl. We believe this is due to the high stability of disordered, random and ordered, intermetallic phases respectively in the CoNiCrFeMn and FeAl systems. This work provides future avenues to the prediction of thermal interstitials and vacancies in solid solutions, which is necessary for models of non-equilibrium behavior of solid solutions.
    \end{abstract}
  \vspace{0.5cm}
  \end{@twocolumnfalse}
]

\section{Introduction}

High entropy alloys (HEAs) have recently attracted interest due to their complex design spaces. In particular, the Cantor alloy (equiatomic CoNiCrFeMn) has been reported to have exceptional mechanical properties\cite{OTTO20135743}  and irradiation resistance\cite{ZHANG2022100807}. However, modeling thermodynamic and kinetic properties of HEAs is not trivial due to their inherent complexities, including large variances in local chemistry, complex thermodynamic behavior, characterization of microstructure, and more. Further complicating the modeling, these individual features can strongly interact with one another. For example, Otto et al. observed the precipitation of a Cr-rich $\sigma$-phase at $\SI{700}{^\circ C}$ and L$1_0$-NiMn, B2-FeCo, and body-centered cubic (bcc) Cr phases at $\SI{500}{^\circ C}$\cite{OTTO201640}, and Li et al. showed that this phase decomposition is driven by NiMn segregation at grain boundaries\cite{10.1063/5.0069107}.

Additionally, the properties of concentrated alloys can be significantly changed by the presence of dilute substitutional and interstitial impurities. For example, in austenitic stainless steels, Muramaki et al. showed that the concentration of octahedral H significantly affects loading damage and fatigue crack growth rate \cite{Murakami2008}, Morgan et al. correlated long-term $^3$H exposure to decreased fracture toughness \cite{doi:10.1080/15361055.2019.1704138}, and Miura et al. observed a decreased fracture stress of grain boundary brittle fracture from increased He content \cite{MIURA2015279}.

Although kinetic properties of concentrated alloys drive non-equilibrium phenomena, the effect of solid solution noise on diffusion is not clear in HEAs. The sluggish diffusion effect is commonly cited as a general property of HEAs\cite{TSAI20134887}, however various studies have found no clear diffusion retardation in many concentrated multicomponent alloys\cite{divinski2018mystery,dkabrowa2019demystifying,daw2021sluggish}. Clarifying this controversy is paramount to understanding diffusion and out-of-equilibrium phenomena in HEAs.

Various out-of-equilibrium phenomena, such as solidification\cite{CAROLI19861867}, dendritic growth\cite{HUNZIKER20014191}, and radiation-induced segregation\cite{NASTAR2012471} are often modeled using linear response theory:

\begin{equation}
    \mathbf{J}_\alpha = -\sum_{\alpha'} L_{\alpha\alpha'} \nabla\mu_{\alpha'}
\end{equation}

where $\mathbf{J}_\alpha$ is the flux of species $\alpha$, $\mu_{\alpha'}$ is the chemical potential of species $\alpha'$, and $L_{\alpha\alpha'}$ is the \textit{equilibrium} $\alpha$-$\alpha'$ Onsager coefficient, which is a transport coefficient coupling species $\alpha$ and $\alpha'$, which additionally obeys the Onsager reciprocal relation $L_{\alpha\alpha'} = L_{\alpha'\alpha}$\cite{PhysRev.38.2265}.

$L_{\alpha\alpha'}$ is usually estimated by inserting one point defect (PD) in a periodic box, simulating transport using an atomistic method such as molecular dynamics \cite{doi:10.1080/08927022.2020.1810685} or atomistic kinetic Monte Carlo\cite{PhysRevB.88.134207,B101982L,PIOCHAUD2016249} and evaluating the generalized Einstein relation\cite{ARAllnatt_1982}:

\begin{equation}
    L_{\alpha\alpha'}^{(1)} = \frac{1}{Vk_BT}\lim_{t\to\infty}\frac{\left\langle \mathbf{R}^{(1)}_\alpha(t)\cdot \mathbf{R}^{(1)}_{\alpha'}(t)\right\rangle}{6t}
\end{equation}

where $V$ is the volume of the box, $k_B$ is the Boltzmann constant, $T$ is temperature, $\mathbf{R}_\alpha^{(1)}(t)$ is the total displacement of $\alpha$ at time $t$ mediated by one PD, $L_{\alpha\alpha'}^{(1)}$ is the $\alpha$-$\alpha'$ transport coefficient mediated by one PD, and $\left\langle\cdot\right\rangle$ denotes an ensemble average.

The use of such computationally expensive techniques introduces a system size constraint, though, which results in a PD concentration much larger than that in thermal equilibrium, significantly overestimating the true equilibrium $L_{\alpha\alpha'}$ necessary for linear response theory. Furthermore, $L_{\alpha\alpha'}^{(1)}$ is inherently size-dependent, while $L_{\alpha\alpha'} = nL_{\alpha\alpha'}^{(1)}$ is intensive, where $n$ is the equilibrium number of defects in the box. Therefore, understanding PD concentration at equilibrium is necessary to accurately model diffusion in concentrated solid solutions.

Traditionally used models for PD concentration in disordered alloys focus on calculating the distribution of PD formation energies by creating a reference configuration and many defective configurations. In the case of thermal vacancies, these defective configurations are created by removing each atom at each lattice site in the reference configuration and calculating their energy differences with the reference configuration using \textit{ab initio} techniques\cite{PhysRevB.93.134115,MANZOOR2021110669,CHEN2018355,PhysRevB.89.024101,10.1063/1.5086172}. Using these models, at each site $\sigma$, there is an occupying type $t(\sigma)$, and one considers the two-state system:

\begin{equation}
    t(\sigma) \text{ at } \sigma \rightleftharpoons \sigma \text{ vacant}
\end{equation}

which denotes local chemical equilibrium at site $\sigma$ between an atom of type $t(\sigma)$ occupying $\sigma$ and no atoms occupying site $\sigma$. From this, one calculates an effective vacancy formation energy from the resulting statistics over lattice sites, treating each site as an independent two-state system.

Various statistical models exist for ordered alloys\cite{PhysRevB.63.094103,PhysRevB.59.6072,MAYER19972207}, which follow the framework of the grand-canonical ensemble, treating chemical potentials as Lagrange multipliers of the numbers of each alloying element.

However, creating a similar statistical model for a random alloy is not as trivial, since other possible configurations are equiprobable by definition. In an approximately random alloy, some chemical ordering will be present, but differing atom types have comparable occupation probabilities at a given site, including PDs, yielding non-zero statistical fluctuations in the local composition at a given lattice site, which necessitate the use of a statistical model that better explores the space of configurations available to the random alloy.

Exploration of this large configuration space can be efficiently achieved through Monte Carlo simulations. For example, Belak and Van der Ven developed a statistical, coarse-grained Monte Carlo method to predict vacancy concentration in disordered alloys\cite{PhysRevB.91.224109}. However, performing Monte Carlo simulations is expensive for large systems, and thus a model only depending on quickly calculable thermodynamic variables is preferable. To explore this configuration space efficiently, we argue that one must also take into account \textit{any} possible atomic type occupying a lattice site.

In this work, we present a Langmuir-like statistical model for competitive occupation at individual lattice sites $\{\sigma\}$. Within this framework, we treat each site as an independent $(k+1)$-state system, where $k$ is the number of alloying elements. From this model, we can calculate an effective formation energy and formation volume for a given impurity, as well as an effective formation quantity for a general thermodynamic force. We then use this model to predict the temperature-dependent vacancy concentration, formation energy, and formation volume for the equiatomic CoNiCrFeMn alloy. To better understand the effects of chemical ordering on vacancy thermodynamics, we additionally perform the same analysis for the equiatomic FeAl alloy because of its tendency to form ordered intermetallic phases \cite{McKamey1996}, in contrast to the CoNiCrFeMn alloy's tendency to form unordered phases.

\section{Model}

\subsection{Statistics of Discrete Systems}

In general, for a system with possible states $\mathcal{F}$, the probability that the system is in state $\phi\in\mathcal{F}$ at absolute temperature $T$ and constant thermodynamic variables $(X_1, X_2, \cdots)$ (where $X_i\neq T$) is:

\begin{equation}
    p_\phi(\beta) = \frac{1}{Z}e^{-\beta G_\phi}
\end{equation}

where $G_\phi$ is the generalized energy for a given statistical ensemble, $\beta = 1/k_BT$ where $k_B = \SI{8.617e-5}{eV\;K^{-1}}$ is the Boltzmann constant, and $Z$ is the partition function:

\begin{equation}
    Z = \sum_{\phi\in\mathcal{F}} e^{-\beta G_\phi}
\end{equation}

which is necessary to fulfill the constraint that $\sum_{\phi\in\mathcal{F}} p_\phi(\beta) = 1$ for all $\beta$. $G_\phi$ depends on the ambient conditions, and is obtained via one or many Legendre transform(s) of the internal energy:

\begin{equation}\label{eq:energy-transform}
    G_\phi = U_\phi - X_1Y_1^{(\phi)} - X_2Y_2^{(\phi)} - \cdots
\end{equation}

where $U_\phi$ is the internal energy of state $\phi$, $Y_1^{(\phi)}$ is the conjugate variable of $X_1$ for state $\phi$, $Y_2^{(\phi)}$ is the conjugate variable of $X_2$ for state $\phi$, and so on. Simplifying yields:

\begin{equation}\label{eq:statistical-prob}
    p_\phi(\beta) = \frac{e^{-\beta G_\phi}}{\displaystyle\sum_{\phi'\in\mathcal{F}} e^{-\beta G_{\phi'}}} = \frac{1}{\displaystyle 1 + \sum_{\phi'\neq\phi} e^{\beta (G_\phi - G_{\phi'})}}
\end{equation}

Langmuir's theory of adsorption, for example, predicts the surface fraction of gas particles adsorbed onto a solid's surface at equilibrium\cite{doi:10.1021/acs.langmuir.9b00154}. This theory can be derived by considering a solid in equilibrium coupled with a gas that can exchange energy and particles, yielding the use of the grand-canonical ensemble to describe the statistical mechanics of the system, and a generalized energy:

\begin{equation}
    G_\phi = U_\phi - \mu N_\phi
\end{equation}

where $U_\phi$ and $N_\phi$ are respectively the internal energy and number of particles in state $\phi$. Then, the state space $\mathcal{F}$ consists of an adsorbed state $(U_\phi, N_\phi) = (E_b, 1)$ and an un-adsorped state $(U_\phi, N_\phi) = (0, 0)$, yielding for the probability of adsorption:

\begin{equation}\label{eq:langmuir}
    p_\text{ads}(\beta) = \frac{1}{1+e^{\beta(E_b - \mu)}}
\end{equation}

where $E_b$ is the binding energy of the gaseous particle onto the surface, and $\mu$ is its chemical potential. Since every surface site is assumed to be identical, $p_\text{ads}(\beta)$ is exactly the surface fraction of gaseous particles adsorbed by the surface.

In this work, we draw an analogy to Langmuir's theory of adsorption by performing a similar discrete system analysis on a solid lattice coupled to a reservoir, exchanging volume, energy, and particles, following the completely open statistical pseudo-ensemble.

From this analysis, we derive expressions for the concentrations of dilute vacancies, substitutional atomic impurities, and small interstitial impurities in an arbitrarily noisy solid solution, and compute vacancy formation energy and volume for CoNiCrFeMn and FeAl as a function of short range order using respective interatomic potentials from Choi et al.\cite{Choi2018} and Lee \& Lee\cite{Lee_2010}.

\subsection{Vacancies}

Consider a lattice with $N$ sites, with each site indexed by an integer $\sigma$. In this lattice, the set of possible occupying atomic types is $\mathcal{A}$, with $N_\alpha$ number of $\alpha$ atoms for all $\alpha\in\mathcal{A}$, and each occupying type of this configuration is denoted by $t(\sigma)$.

To fully sample energetic states on this lattice, one would have to sample every possible configuration. However, this is impractical for a large system, with the number of states being $\mathcal{O}\left(\|\mathcal{A}\|^N\right)$ for an equiatomic alloy. Instead, we assume that we \textit{a priori} know a stable initial state, and sample states near that stable state by swapping atoms in that configuration with atoms of differing chemical species. This configuration must be stable in the sense that timescales are well-separated, such that the configuration can be considered to be in a quasi-equilibrium that we can sample with equilibrium statistical mechanics. However, note that such a configuration could be resampled during the data collection process and conceptually it is not a fixed configuration to which energies are referred.

To do this, couple the lattice with an energy, volume, and atom-exchanging reservoir with types  $\mathcal{A}_r = \mathcal{A}\cup\{v\}$, where $v$ is a vacancy. The reservoir then controls the chemical potentials of $\mathcal{A}_r$ as well as the pressure and temperature of the lattice, thus the probability that a site $\sigma$ is occupied by any type in $\mathcal{A}_r$ follows the completely open ensemble, or equivalently the $\mu pT$ ensemble\cite{Marzolino2021}. For any given $\alpha\in\mathcal{A}$, the probability that $\alpha$ occupies site $\sigma$ is then:

\begin{equation}
    p_{(\sigma, \alpha)}(\beta) = \frac{e^{-\beta\left(E_\sigma^{(\alpha)} + pV_\sigma^{(\alpha)} - \sum_{\alpha'} \mu_{\alpha'} N_{\alpha'} + \mu_{t(\sigma)} - \mu_{\alpha}\right)}}{Z_\sigma(\beta)}
\end{equation}

where $E_\sigma^{(\alpha)}$ and $V_\sigma^{(\alpha)}$ are respectively the energy and volume of the lattice when type $\alpha$ occupies site $\sigma$, and $p$ is the pressure of the reservoir, and $Z_\sigma(\beta)$ is the partition function, constraining that the probabilities of all available states sum to unity:

\begin{equation}
    \begin{aligned}
        Z_\sigma(\beta) &= e^{-\beta \left( \varepsilon_\sigma + pv_\sigma - \sum_{\alpha'} \mu_{\alpha'} N_{\alpha'} + \mu_{t(\sigma)}\right)}\\
        &+ \sum_\alpha e^{-\beta\left(E_\sigma^{(\alpha)} + pV_\sigma^{(\alpha)} - \sum_{\alpha'} \mu_{\alpha'} N_{\alpha'} + \mu_{t(\sigma)} - \mu_{\alpha}\right)}
    \end{aligned}
\end{equation}

Here, $\alpha'$ is merely a dummy index, denoting another sum over all atom types $\alpha'\in\mathcal{A}$. In this work, we compute these energies and volumes from molecular statics with an interatomic potential, but these quantities can also be computed with various atomistic methods, including but not limited to density functional theory \cite{Giustino_2020} and/or a cluster expansion \cite{WU2016243}.

Similarly, the probability that $v$ occupies site $\sigma$ is then:

\begin{equation}
    \begin{aligned}
        p_{(\sigma, v)}(\beta) &= \frac{e^{-\beta \left( \varepsilon_\sigma + pv_\sigma - \sum_{\alpha'} \mu_{\alpha'} N_{\alpha'} + \mu_{t(\sigma)} \right)}}{Z_\sigma(\beta)}
    \end{aligned}
\end{equation}

where $\varepsilon_\sigma$ and $v_\sigma$ are respectively the energy and volume of the lattice when the vacancy type $v$ occupies site $\sigma$. The probability that $v$ occupies site $\sigma$ is then:

\begin{equation}\label{eq:local-prob}
    \begin{aligned}
        p_{(\sigma, v)}(\beta)
        &=\frac{1}{1+\sum_{\alpha}e^{\beta\mathcal{H}_\sigma^{(\alpha)}}}
    \end{aligned}
\end{equation}

where  $\mathcal{H}_\sigma^{(\alpha)}$ is the effective enthalpic penalty of swapping an $\alpha$ atom for the vacancy type $v$:

\begin{equation}\label{eq:enthalpic_penalty}
    \mathcal{H}_\sigma^{(\alpha)} = \varepsilon_\sigma - E_\sigma^{(\alpha)} + p\left(v_\sigma - V_\sigma^{(\alpha)}\right) + \mu_\alpha
\end{equation}

matching Eq.~\eqref{eq:statistical-prob}, summing over $\alpha\in\mathcal{A}_r\setminus \{v\} = \mathcal{A}$. Note that the resulting expression for the local occupation probability is analogous to the expression for Langmuir adsorption (Eq.~\eqref{eq:langmuir}). A notable difference, however, is that we expect vacancies to be energetically unfavorable, while the binding energy in the Langmuir model is negative, making binding favorable. This results in opposite trends in temperature, i.e. surface fraction in adsorption decreases with temperature, while vacancy concentration increases with temperature.

From the local $v$-occupation probabilities, we can calculate the global concentration of $v$ by averaging the $v$-occupation probability over sites $\{\sigma\}$:

\begin{equation}\label{eq:concentration}
    x_v(\beta) = \left\langle p_{(\sigma, v)}(\beta)\right\rangle
\end{equation}

This analysis is equivalent to factorizing the total partition function into single-site contributions:

\begin{equation}
    Z_\text{lattice}(\beta) = \prod_\sigma Z_\sigma(\beta)
\end{equation}

where $Z_\text{lattice}$ is the $\mu pT$ partition function for the entire lattice. In the case of strongly interacting sites $\sigma$, this factorization may not be accurate, necessitating the inclusions of correlations between $\sigma$ and its neighbors. For the purposes of dilute impurities, however, we assume that we can neglect correlations between neighboring sites.

Note that this model is quickly adaptable to the more well-established grand canonical ensemble. In this case, we instead have:

\begin{equation}
    \mathcal{H}_\sigma^{(\alpha)} = \varepsilon_\sigma - E_\sigma^{(\alpha)} + \mu_\alpha
\end{equation}

which is intuitively equivalent to \eqref{eq:enthalpic_penalty} if the $p\Delta V$ term is small. It is worth noting that the $\mu pT$ ensemble is controversial and often considered ambiguous and non-rigorous, since, for large systems, its corresponding thermodynamic potential is identically $0$ and $\mu$, $p$, and $T$ cannot be varied independently\cite{10.1063/1.1732447}. However, we will use the completely open ensemble for the remainder of the calculations, since this ensemble allows for the computation of impurity formation volumes. This can be briefly justified by noting that vacancy formation volumes tend to be on the order of $\SI{100}{bohrs^3}\sim \SI{10}{\AA^3}$\cite{PhysRevB.59.11693}, yielding a $p\Delta V$ term on the order of $\SI{1e-5}{eV}$ at $p = \SI{1}{bar}$, and can therefore be considered negligible when comparing to calculations using the grand canonical ensemble.

\subsection{Substitutional atom}

For substitutional atoms, the only difference in the model opposed to vacancies is that an atomic impurity has a non-zero chemical potential. The probability that an atom type $\iota$ occupies a site $\sigma$ is:

\begin{equation}
    \begin{aligned}
        p_{(\sigma, \iota)}(\beta) &= \frac{e^{-\beta \left( \varepsilon_\sigma + pv_\sigma - \sum_{\alpha'} \mu_{\alpha'} N_{\alpha'} + \mu_{t(\sigma)} - \mu_\iota \right)}}{Z_\sigma(\beta)}
    \end{aligned}
\end{equation}

with local partition function $Z_\sigma(\beta)$:

\begin{equation}
    \begin{aligned}
        Z_\sigma(\beta) &= e^{-\beta \left( \varepsilon_\sigma + pv_\sigma - \sum_{\alpha'} \mu_{\alpha'} N_{\alpha'} + \mu_{t(\sigma)} - \mu_\iota\right)}\\
                    &+ \sum_\alpha e^{-\beta\left(E_\sigma^{(\alpha)} + pV_\sigma^{(\alpha)} - \sum_{\alpha'} \mu_{\alpha'} N_{\alpha'} + \mu_{t(\sigma)} - \mu_{\alpha}\right)}
    \end{aligned}
\end{equation}

therefore, the occupation probability is:

\begin{equation}
    p_{(\sigma, \iota)}(\beta) = \frac{1}{1+ \sum_\alpha e^{\beta \left( \varepsilon_\sigma -E_\sigma^{(\alpha)} + p\left(v_\sigma - V_\sigma^{(\alpha)}\right) + \mu_{\alpha} - \mu_\iota\right) }}
\end{equation}

which is equivalent to Eq.~\eqref{eq:local-prob}, with a new expression for $\mathcal{H}_\sigma^{(\alpha)}$:

\begin{equation}
    \mathcal{H}_\sigma^{(\alpha)} = \varepsilon_\sigma - E_\sigma^{(\alpha)} + p\left(v_\sigma - V_\sigma^{(\alpha)}\right) + \mu_\alpha - \mu_\iota
\end{equation}

\subsection{Small interstitial atom}

Sufficiently small elements will occupy interstitial sites rather than lattice sites, so we instead track the probability that an interstitial site $\sigma$ is occupied by a small atom type $\iota$ from a set of atom types $\mathcal{I}$. For tetrahedral, octahedral, and cubic interstitial sites, the maximum radius ratio ($\text{radius of impurity} / \text{radius of host}$) for an atom to be sufficiently small is roughly $0.225$, $0.414$, and $0.154$\cite{Shackelford_2016}. The probability that a small impurity $\iota\in\mathcal{I}$ occupies a site $\sigma$ is then:

\begin{equation}
    \begin{aligned}
        p_{(\sigma, \iota)}(\beta) &= \frac{e^{-\beta\left(E_\sigma^{(\iota)} + pV_\sigma^{(\iota)} - \sum_{\alpha'}\mu_{\alpha'}N_{\alpha'} - \mu_\iota\right)}}{Z_\sigma(\beta)}
    \end{aligned}
\end{equation}

with local partition function $Z_\sigma(\beta)$:

\begin{equation}
    \begin{aligned}
        Z_\sigma(\beta) &= e^{-\beta\left(E^\circ + pV^\circ - \sum_{\alpha'}\mu_{\alpha'}N_{\alpha'}\right)}\\
        &+\sum_{\iota'} e^{-\beta\left(E_\sigma^{(\iota')} + pV_\sigma^{(\iota')} - \sum_{\alpha'}\mu_{\alpha'}N_{\alpha'} - \mu_{\iota'}\right)}
    \end{aligned}
\end{equation}

Therefore:

\begin{equation}
    \begin{aligned}
        p_{(\sigma, \iota)}(\beta) &= \frac{1}{1 + e^{ \beta\mathcal{H}_{\sigma}^{(\iota)} } +  \sum_{\iota'\neq \iota} e^{\beta\mathcal{H}_\sigma^{(\iota, \iota')}}}
    \end{aligned}
\end{equation}

where we have two enthalpic penalties, one for placing an atom of type $\iota$ at interstitial site $\sigma$:

\begin{equation}
    \mathcal{H}_{\sigma}^{(\iota)} = E_\sigma^{(\iota)} - E^\circ + p\left(V_\sigma^{(\iota)} - V^\circ\right) - \mu_\iota
\end{equation}

and one for swapping an atom of type $\iota'$ with an atom of type $\iota$ at interstitial site $\sigma$:

\begin{equation}
    \mathcal{H}_\sigma^{(\iota, \iota')} = E_\sigma^{(\iota)} - E_\sigma^{(\iota')} + p\left(V_\sigma^{(\iota)} - V_\sigma^{(\iota')}\right) - (\mu_\iota - \mu_{\iota'})
\end{equation}

where $E_\sigma^{(\iota)}$ and $V_\sigma^{(\iota)}$ are respectively the relaxed energy and volume of the lattice with an atom $\iota$ at interstitial site $\sigma$, and $E^\circ$ and $V^\circ$ are respectively the relaxed energy and volume of the lattice without extra atoms in interstitial sites.

Here, we count $|\mathcal{I}| + 1$ states per site $\sigma$, where $|\mathcal{I}|$ is the number of possible occupying atom types, assuming that the larger $\mathcal{A}$ atoms are not stable at interstitial sites. Then, the fraction of interstitial sites occupied by $\iota$ is $x_\iota(\beta) = \left\langle p_{(\sigma, \iota)}(\beta)\right\rangle$.

Note that, in the homogeneous limit for a single atom type, we are left with a single enthalpic penalty $\mathcal{H}$, recovering the standard formation expression $x(\beta) = e^{-\beta\mathcal{H}}$ for substitutional atoms, vacancies, and small interstitial elements. Additionally, in the high-temperature limit, we recover an effective enthalpic penalty:

\begin{equation}
    \mathcal{H} = \left\langle \frac{1}{|\mathcal{S}|}\sum_{s\in \mathcal{S}}\mathcal{H}_\sigma^{(s)}\right\rangle
\end{equation}

where $\mathcal{S} = \mathcal{A}$ for substitutional atoms and vacancies, and $\mathcal{S} = \{\iota\}\cup\{(\iota, \iota')\;|\;\iota'\in\mathcal{I} \text{ and } \iota'\neq\iota\}$ for small interstitial atoms (see \ref{limits}).
\section{Methods}

To evaluate the concentration of a defect in a model concentrated alloy, we chose vacancies in equiatomic, face-centered cubic (fcc) CoNiCrFeMn and body-centered cubic (bcc) FeAl alloys, performing atomistic calculations using the Large-scale Atomic/Molecular Massively Parallel Simulator (LAMMPS)\cite{THOMPSON2022108171,lammps}. For the CoNiCrFeMn alloy, we use a modified embedded-atom method\cite{PhysRevB.40.6085} (MEAM) potential developed by Choi et al. to study the effects of individual elements on solid solution hardening \cite{Choi2018}. For FeAl, we use a MEAM potential developed by Lee and Lee to study defect formation behavior \cite{Lee_2010}.

\subsection{Equilibration}

Using LAMMPS, we first initialize a random $7\times 7\times 7$ fcc CoNiCrFeMn alloy with lattice parameter $\SI{3}{\AA}$ and minimize the configuration at $\SI{0}{bar}$ \cite{10.1063/1.328693}. Then, we perform an NPT equilibration at $\SI{600}{K}$ and $\SI{1}{bar}$ for $10^4$ steps with a femtosecond timestep using an integration scheme proposed by Shinoda et al.\cite{PhysRevB.69.134103} with a temperature dampening time of $\SI{0.1}{ps}$ and a pressure dampening time of $\SI{1.0}{ps}$. Lastly, we run a hybrid canonical Monte Carlo/NPT molecular dynamics (MC-MD) routine, where we attempt five $\alpha$-$\alpha'$ swaps per $\alpha$-$\alpha'$ pair every $10^3$ MD steps for $10^6$ total MD steps at $\SI{600}{K}$, accepting or rejecting each swap with the Metropolis criterion \cite{10.1063/1.1699114} using the MC package in LAMMPS\cite{PhysRevB.85.184203}.

\begin{figure}[H]
    \centering
    \includegraphics[width=\linewidth]{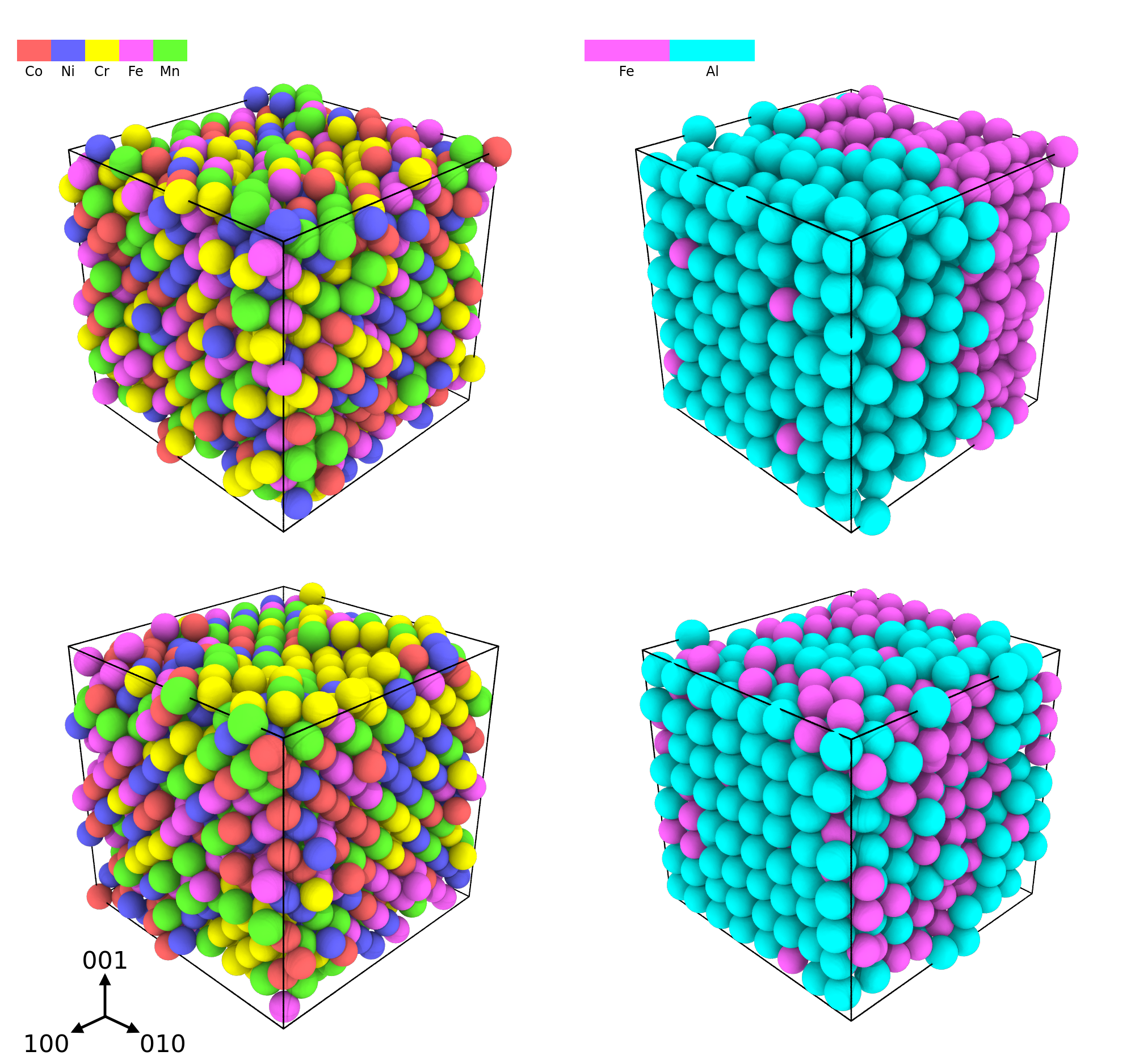}
    \caption{MC-MD snapshots of the CoNiCrFeMn (left) and FeAl (right) systems at $\SI{0}{ns}$ (top) and $\SI{1}{ns}$ (bottom) rendered with OVITO\cite{ovito}.}
    \label{fig:mc-md}
\end{figure}

We perform a similar procedure for FeAl, instead starting with a segregated $8\times 8\times 8$ bcc solution, and attempting $25$ $\alpha$-$\alpha'$ swaps every $10^3$ steps (see Figure \ref{fig:mc-md}).

\begin{figure}[H]
    \centering
    \includegraphics[width=\linewidth]{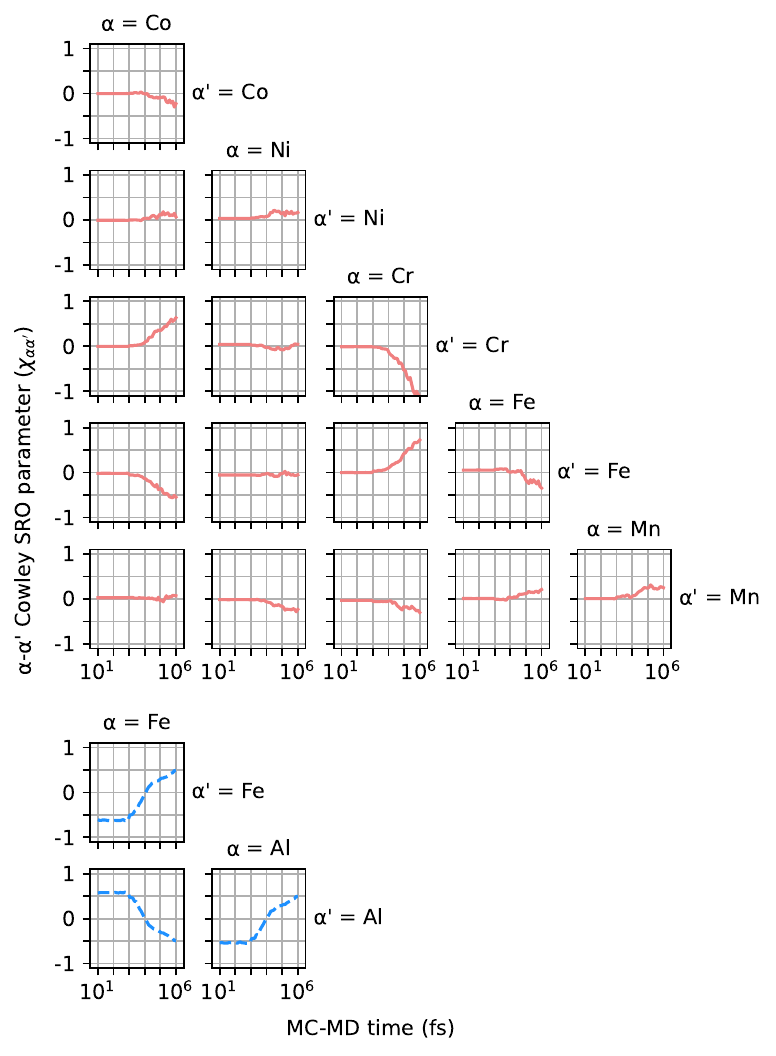}
    \caption{Cowley SRO parameters $\chi_{\alpha\alpha'}$ as a function of MC-MD simulation time for the CoNiCrFeMn (top) and FeAl (bottom) systems.}
    \label{fig:sro_params}
\end{figure}

In the annealing process shown in Figure \ref{fig:mc-md}, there is clear chemical ordering, most notably Cr segregation in CoNiCrFeMn and the formation of a B2 phase in FeAl. To quantify the equilibration, we \sout{first} calculated the Cowley short-range order (SRO) parameter\cite{PhysRev.77.669} for each $\alpha$-$\alpha'$ type pair at each timestep in our MC-MD routine, shown in Fig.~\ref{fig:sro_params}:

\begin{equation}\label{eq:sro}
    \chi_{\alpha\alpha'} = 1 - \frac{p_{\alpha\alpha'}}{(2 - \delta_{\alpha\alpha'})x_\alpha x_{\alpha'}}
\end{equation}

where $p_{\alpha\alpha'}$ is the probability of finding types $\alpha$ and $\alpha'$ in the first nearest-neighbor shell and $\delta_{\alpha\alpha'}$ is the Kronecker delta:

\begin{equation}
    \delta_{\alpha\alpha'} = \begin{cases} 1 & \alpha = \alpha' \\ 0 & \text{else} \end{cases}
\end{equation}

$\chi_{\alpha\alpha'}$ therefore describes the tendency for $\alpha$ atoms to congregate near $\alpha'$ atoms in the alloy. Respectively, a positive and negative $\chi_{\alpha\alpha'}$ indicates that $\alpha$ and $\alpha'$ atoms are unlikely and likely to be neighbors in the solution, while a zero $\chi_{\alpha\alpha'}$ indicates that the positions of $\alpha$ and $\alpha'$ atoms are uncorrelated. For a perfectly random solution, $\chi_{\alpha\alpha'} = 0$ for all $\alpha$, $\alpha'$ pairs.

We note that $\chi_{\text{Cr}\text{Cr}}$ approaches $-1$ in CoNiCrFeMn, denoting a strong tendency to self-segregate. Similarly, $\chi_{\text{Fe}\text{Al}}$ approaches $-0.5$ in FeAl, denoting a tendency to form an intermetallic phase. Asymptotic behavior in these SRO time series corresponds to thermodynamic equilibration, approaching the equilibrium SROs as the MC-MD time goes to infinity. Here, we note that the configurations are not fully equilibrated, since our focus is on vacancy thermodynamics as we vary chemical ordering.

Note also that the MC-MD steps are sampled at a logarithmic frequency (50 steps between $\SI{10}{fs}$ and $\SI{1}{ns}$) in order to obtain high resolution information over the fast initial equilibration steps. Since we are performing MC swaps, it is additionally important to note that the MC-MD time is not physical, but rather measures the amount of MC steps attempted.

Although Eq.~\eqref{eq:sro} differs from the traditional expression for the Cowley SRO parameters, which uses a conditional probability as opposed to an unconditional probability, Rao and Curtin proved that they are equivalent\cite{RAO2022117621}. The $\chi_{\alpha\alpha'}$ calculations are implemented as an OVITO modifier in a lightweight library available in the Python Package Index\cite{sro_pypi}.

\subsection{Energetics Calculation}

After equilibrating with MC-MD, the resulting configuration is relaxed at $\SI{0}{bar}$.

Then, we select the site $\sigma = 1$ (starting site indexing at $1$), inserting all types $\alpha\in\mathcal{A}$ and again minimizing at $\SI{0}{bar}$, calculating the resulting interaction energy $E_\sigma^{(\alpha)}$ and volume $V_\sigma^{(\alpha)}$. For the same lattice site, we then remove the current occupying atom, once again minimizing at $\SI{0}{bar}$ and calculating the resulting interaction energy $\varepsilon_\sigma$ and volume $v_\sigma$.

Then, we iterate to the next site $\sigma\leftarrow \sigma + 1$, loading in the aforementioned $\SI{0}{bar}$ configuration, and repeating the aforementioned energetics/volumetrics calculations, storing all values $E_\sigma^{(\alpha)}$, $V_\sigma^{(\alpha)}$, $\varepsilon_\sigma$, and $v_\sigma$ for post-processing.

\subsection{Chemical Potential Calculation}

After adding an atom of type $\alpha$ to a system with free energy $G^\circ$, the new free energy is:

\begin{equation}
    G_\alpha = G^\circ + \mu_\alpha
\end{equation}

Therefore, swapping an $\alpha$ atom with an $\alpha'$ atom results in a free energy change equal to their respective chemical potential differences:

\begin{equation}\label{eq:free_energy_difference}
    G_\alpha - G_{\alpha'} = \mu_\alpha - \mu_{\alpha'}
\end{equation}

As an approximation, we evaluate free energy differences at $\SI{0}{K}$ and $\SI{0}{bar}$ for each $\alpha$-$\alpha'$ pair, where free energy is equal to internal energy:

\begin{equation}\label{eq:energy_difference}
    \left\langle E_\sigma^{(\alpha)} - E_\sigma^{(\alpha')}\right\rangle = \mu_\alpha - \mu_{\alpha'}
\end{equation}

Additionally, using the Euler equation for enthalpy at $\SI{0}{K}$ and $\SI{0}{bar}$ yields:

\begin{equation}\label{eq:euler_eq}
    h = \sum_\alpha x_\alpha \mu_\alpha
\end{equation}

where $h$ is the enthalpy per atom of the well-equilibrated crystal. For a $k$-component system, we can generate $m = \frac{1}{2}k(k-1)$ equations from \eqref{eq:energy_difference} (one per $(\alpha, \alpha')$ pair), plus \eqref{eq:euler_eq}, which is an overdetermined system for $k > 2$, defined by:

\begin{equation}\label{eq:linear-sys}
    \boldsymbol{A}\boldsymbol{\mu} = \boldsymbol{\Delta}
\end{equation}

where, if each $(\alpha, \alpha')$ pair is indexed by $0\leq i < m$, then the elements of $\boldsymbol{A}$, $\boldsymbol{\mu}$, and $\boldsymbol{\Delta}$ are defined by:

\begin{equation}\label{eq:linear-sys2}
    \begin{aligned}
        A_{i\alpha} &= 1\\
        A_{i\alpha'} &= -1\\
        A_{m\alpha} &= x_\alpha \\
        \Delta_i &= \left\langle E_\sigma^{(\alpha)} - E_\sigma^{(\alpha')}\right\rangle\\
        \Delta_m &= h \\
        \mu_\alpha &= \mu_\alpha
    \end{aligned}
\end{equation}

\begin{table}[H]
    \begin{tabular}{|c||c|c|c|c|c|}
        \hline
        $\alpha$ & Co & Ni & Cr & Fe & Mn \\
        \hline
        $\mu_\alpha$ & $-4.48$ & $-4.57$ & $-3.95$ & $-4.28$ & $-3.09$ \\
        \hline
    \end{tabular}

    \vspace{0.5cm}

    \begin{tabular}{|c||c|c|}
        \hline
        $\alpha$ & Fe & Al \\
        \hline
        $\mu_\alpha$ & $-4.62$ & $-3.50$ \\
        \hline
    \end{tabular}
    
    \caption{Chemical potentials (in eV) calculated from \eqref{eq:linear-sys} and \eqref{eq:linear-sys2} at the final MC-MD snapshot for the CoNiCrFeMn (top) and FeAl (bottom) systems.}
    \label{tab:chem-pot}
\end{table}

In an infinite system, where there is zero uncertainty in the average in Eq.~\eqref{eq:energy_difference}, we would not have an overdetermined system. For example, in a $3$ component system, we have the additional equation $\mu_1 - \mu_2 + \mu_2 - \mu_3 = \mu_1 - \mu_3$, meaning the set of equations in Eq.~\eqref{eq:energy_difference} relating the $1$-$2$, $1$-$3$, and $2$-$3$ swaps are linearly dependent, yielding a determined system with $3$ equations rather than $4$.

In contrast, in a finite system with imperfect statistics, finite uncertainty will drift the mean energy difference from these swaps, yielding linearly independent equations. Such equations must be solved approximately, usually in the form of a least-squares solution that finds the solution $\boldsymbol{\mu}$ which minimizes the error $\|\boldsymbol{A}\boldsymbol{\mu} - \boldsymbol{\Delta}\|$. In Table \ref{tab:chem-pot}, these errors are respectively $\SI{4.8e-15}{eV}$ and $\SI{8.9e-16}{eV}$ in CoNiCrFeMn and FeAl, indicating a very accurate fit.

Note that Eq.~\eqref{eq:energy_difference} closely resembles the Widom insertion method\cite{10.1063/1.1734110}. The traditional implementation of this method, where single particles are removed, implicitly adds vacancies to the system, which is unnecessary for calculating chemical potentials, since chemical potentials of the alloying elements should be independent of the presence of point defects. Here, we instead compute chemical potentials from the occupation energies $E_\sigma^{(\alpha)}$ for the alloying types $\alpha\in\mathcal{A}$, not including vacancies, which removes the need to correct for the addition of a high-energy vacancy.

Instead of relying on a $\SI{0}{K}$ and $\SI{0}{bar}$ approximation for use in Eq.~\eqref{eq:free_energy_difference}, one can also calculate free energy at a given temperature and pressure using thermodynamic integration methods (such as adiabatic switching \cite{PhysRevE.53.465} \cite{FREITAS2016333}, finite difference thermodynamic integration \cite{10.1063/1.452357}, Bennet's acceptance ratio method\cite{BENNETT1976245}, etc.).
\section{Results}

\subsection{Formation Quantities}

Expanding $\ln x_v = \ln x_v(\beta, \beta p)$ about $\beta = \beta^\circ$ and $\beta p = \beta^\circ p^\circ$ to first order in $\beta$ and $\beta p$ gives:

\begin{equation}
    \begin{aligned}
        \ln x_v(\beta, \beta p) &= \ln x_v(\beta^\circ, \beta^\circ p^\circ)\\
        &- (\beta - \beta^\circ) E_\text{form}(\beta^\circ, \beta^\circ p^\circ)\\
        &- (\beta p - \beta^\circ p^\circ) \Omega_\text{form}(\beta^\circ, \beta^\circ p^\circ)
    \end{aligned}
\end{equation}

where $E_\text{form}(\beta^\circ, \beta^\circ p^\circ)$ and $\Omega_\text{form}(\beta^\circ, \beta^\circ p^\circ)$ are respectively the formation energy and formation volume at $(\beta, \beta p) = (\beta^\circ, \beta^\circ p^\circ)$. Therefore:

\begin{equation}
    \begin{aligned}
        E_\text{form}(\beta, \beta p) &= -\left(\frac{\partial\ln x_v(\beta, \beta p)}{\partial \beta}\right)_{\beta p} \\
        &=\frac{1}{x_v(\beta)}\left\langle \frac{\sum_\alpha \mathcal{H}_\sigma^{(\alpha)}e^{\beta\mathcal{H}_\sigma^{(\alpha)}}}{\left(1+\sum_\alpha e^{\beta\mathcal{H}_\sigma^{(\alpha)}}\right)^2}\right\rangle
    \end{aligned}
\end{equation}

and:

\begin{equation}
    \begin{aligned}
        \Omega_\text{form}(\beta, \beta p) &= -\left(\frac{\partial \ln x_v(\beta, \beta p)}{\partial (\beta p)}\right)_\beta \\
        &=\frac{1}{x_v(\beta)}\left\langle \frac{\sum_\alpha \left(v_\sigma - V_\sigma^{(\alpha)}\right)e^{\beta\mathcal{H}_\sigma^{(\alpha)}}}{\left(1+\sum_\alpha e^{\beta\mathcal{H}_\sigma^{(\alpha)}}\right)^2}\right\rangle
    \end{aligned}
\end{equation}

Using our site statistics and the chemical potentials calculated from Eqs.~\eqref{eq:linear-sys} and \eqref{eq:linear-sys2}, we can calculate all $\mathcal{H}_\sigma^{(\alpha)}$'s from Eq.~\eqref{eq:enthalpic_penalty}.

\begin{figure}[H]
    \centering
    \includegraphics[width=\linewidth]{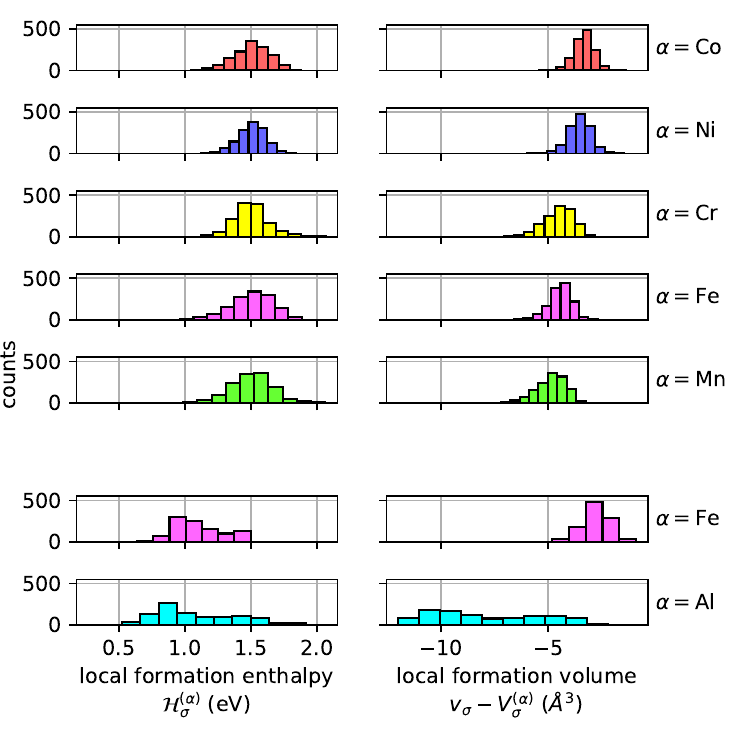}
    \caption{Histograms of local enthalpic and volumetric penalties, respectively $\mathcal{H}_\sigma^{(\alpha)}$ and $v_\sigma - V_\sigma^{(\alpha)}$, at the final MC-MD snapshot for the CoNiCrFeMn (top) and FeAl (bottom) systems using the $(k+1)$-state model.}
    \label{fig:distributions}
\end{figure}

Notably, both formation quantities respectively depend on statistics of enthalpic and volumetric differences between vacant states and $\alpha$-occupying states. Statistics for these quantities for the final MC-MD snapshot are shown in Figure \ref{fig:distributions}.

For CoNiCrFeMn, the distributions of local enthalpic and volumetric penalties are unimodal, which is indicative of a random solution, since the average environment in the solution does not vary spatially. For FeAl, however, the same distributions are more complex. In particular, the distributions of $\mathcal{H}_\sigma^{(\text{Fe})}$, $\mathcal{H}_\sigma^{(\text{Al})}$, and $v_\sigma - V_\sigma^{(\text{Al})}$ are bimodal, with each peak representing the two sublattices in the B2 phase. Using the supercell method\cite{PhysRevB.59.11693} and the chemical potentials in Table \ref{tab:chem-pot}, the formation energies on the Fe and Al sublattices are respectively $\SI{1.40}{eV}$ and $\SI{1.54}{eV}$ (see \ref{supercell}), consistent with the rightmost peaks in the histograms for $\mathcal{H}_\sigma^{(\text{Fe})}$ and $\mathcal{H}_\sigma^{(\text{Al})}$ (see Figure \ref{fig:distributions}). Here, these formation energies correspond to the rightmost peaks since, by definition, the Fe atoms are lower energy on the Fe sublattice, so the energy difference between a vacancy and an Fe atom on the Fe sublattice is larger in magnitude than the energy difference between a vacancy and an Al atom on the Fe sublattice. 

Additionally, the leftmost peaks are shifted by the anti-site defect energies. For $\mathcal{H}_\sigma^{(\text{Fe})}$, this shift is nearly the Al-sublattice anti-site formation energy $\SI{0.51}{eV}$. For $\mathcal{H}_\sigma^{(\text{Al})}$, this shift is close to the Fe-sublattice anti-site formation energy $\SI{0.66}{eV}$.

The intermediate probability density between the peaks denotes that the sample is not fully equilibrated, which is consistent with the prior SRO calculations (see Figure \ref{fig:sro_params}), where the final $\chi_{\text{Fe}\text{Al}} < 1$. From these statistics, we can calculate $x_\text{V}(\beta)$, $E_\text{form}(\beta)$, and $\Omega_\text{form}(\beta)$ for the final MC-MD snapshot. Note that we implicitly set $p = 0$. We can also use these statistics to compute these quantities for the aforementioned traditional two-state model\cite{PhysRevB.93.134115,MANZOOR2021110669,CHEN2018355,PhysRevB.89.024101,10.1063/1.5086172}, which excludes the probability that site $\sigma$ can be occupied by a type other than $t(\sigma)$:

\begin{equation}
    \begin{aligned}
        x_v^\text{2s}(\beta, \beta p) &= \left\langle \frac{1}{1+e^{\beta \mathcal{H}_\sigma^{(t(\sigma))}}}\right\rangle \\
        E_\text{form}^\text{2s}(\beta, \beta p) &= \frac{1}{x_v^\text{2s}(\beta)}\left\langle \frac{ \mathcal{H}_\sigma^{(t(\sigma))}e^{\beta\mathcal{H}_\sigma^{(t(\sigma))}}}{\left(1+ e^{\beta\mathcal{H}_\sigma^{(t(\sigma))}}\right)^2}\right\rangle \\
        \Omega_\text{form}^\text{2s}(\beta, \beta p) &= \frac{1}{x_v^\text{2s}(\beta)}\left\langle \frac{\left(v_\sigma - V_\sigma^{(t(\sigma))}\right)e^{\beta\mathcal{H}_\sigma^{(t(\sigma))}}}{\left(1+ e^{\beta\mathcal{H}_\sigma^{(t(\sigma))}}\right)^2}\right\rangle
    \end{aligned}
\end{equation}

\begin{figure}[H]
    \centering
    \includegraphics[width=\linewidth]{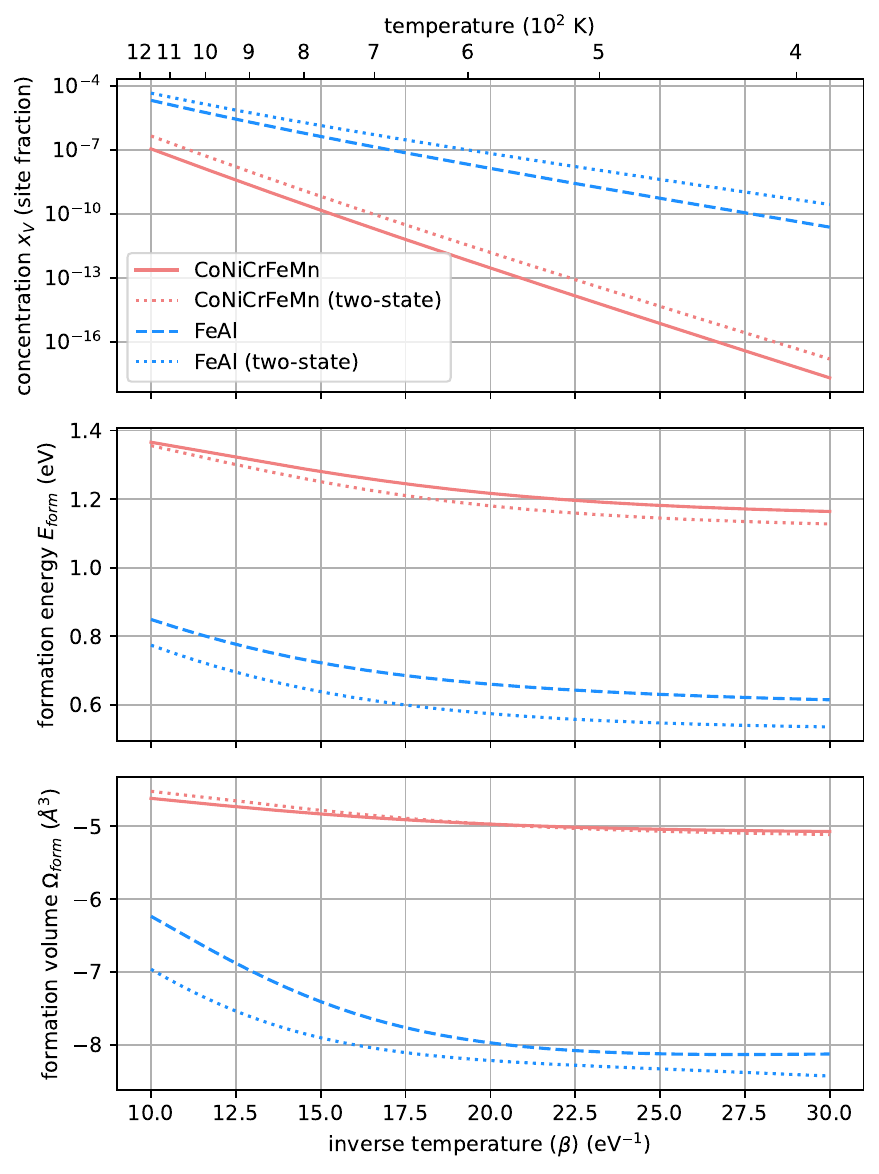}
    \caption{Vacancy concentration (top), formation energy (middle), and formation volume (bottom) as a function of temperature at the final MC-MD snapshot using both the two-state and $(k+1)$-state models.}
    \label{fig:formations}
\end{figure}

Note the strong temperature dependence on the formation energy and formation volume for both the $(k+1)$-state model and the two-state model. Both formation energy and volume monotonically increase with temperature, with formation energy notably changing more than $\SI{0.2}{eV}$ over the $\SI{387}{K}$-$\SI{1161}{K}$ range for both alloy systems.

For CoNiCrFeMn, the $(k+1)$-state model and the two-state model yield similar vacancy concentrations, formation energies, and formation volumes over the tested temperature range. However, the predictions are less similar for FeAl. For FeAl, the difference in predicted formation energy is roughly $\SI{0.08}{eV}$ over the entire tested temperature range, and the difference in predicted formation volume increases from $\SI{0.20}{\AA^3}$ to $\SI{0.73}{\AA^3}$.

More generally, we can calculate any vacancy formation quantity $\xi_\text{form}$ conjugate to some external generalized force $X$ (for example, an external magnetic field) by defining the corresponding formation quantity via an expansion about $(\beta, \beta p, \beta X) = (\beta^\circ, \beta^\circ p^\circ, \beta^\circ X^\circ)$:

\begin{equation}
    \begin{aligned}
        \ln x_v(\beta, \beta p, \beta X) &= \ln x_v(\beta^\circ, \beta^\circ p^\circ, \beta^\circ X^\circ)\\
        &- (\beta - \beta^\circ) E_\text{form}(\beta^\circ, p^\circ, X^\circ)\\
        &-(\beta p - \beta^\circ p^\circ)\Omega_\text{form}(\beta^\circ, \beta^\circ p^\circ, \beta^\circ X^\circ)\\
        &+ (\beta X - \beta^\circ X^\circ)\xi_\text{form}(\beta^\circ, \beta^\circ p^\circ, \beta^\circ X^\circ)
    \end{aligned}
\end{equation}

which then implies:

\begin{equation}
    \begin{aligned}
        \xi_\text{form}(\beta, \beta p, \beta X) &= \left(\frac{\partial \ln x_v(\beta, \beta p, \beta X)}{\partial(\beta X)}\right)_{\beta, \beta p}\\
        &=-\frac{1}{x_v(\beta)}\left\langle \frac{\sum_\alpha \left(\xi_\sigma - \Xi_\sigma^{(\alpha)}\right)e^{\beta\mathcal{H}_\sigma^{(\alpha)}}}{\left(1+\sum_\alpha e^{\beta\mathcal{H}_\sigma^{(\alpha)}}\right)^2}\right\rangle
    \end{aligned}
\end{equation}

where $\xi_\sigma$ is the generalized displacement of the defect occupying site $\sigma$, and $\Xi_\sigma^{(\alpha)}$ is the generalized displacement of an $\alpha$ atom occupying site $\sigma$. In this case, one must additionally modify $\mathcal{H}_\sigma^{(\alpha)}$ to include the new generalized force. For example, the enthalpic penalty for replacing an atom of type $\alpha$ at site $\sigma$ under constant volume, temperature, and a constant, small external magnetic field is:

\begin{equation}
    \mathcal{H}_\sigma^{(\alpha)} = \varepsilon_\sigma - E_\sigma^{(\alpha)} - \mathbf{B}\cdot\left(\mathbf{m} - \mathbf{M}_\sigma^{(\alpha)}\right) + \mu_\alpha
\end{equation}

yielding a magnetic moment of formation:

\begin{equation}
    \mathbf{m}_\text{form}(\beta, \beta\mathbf{B}) = \frac{1}{x_v(\beta)}\left\langle \frac{\sum_\alpha \left(\mathbf{m}_\sigma - \mathbf{M}_\sigma^{(\alpha)}\right)e^{\beta\mathcal{H}_\sigma^{(\alpha)}}}{\left(1+\sum_\alpha e^{\beta\mathcal{H}_\sigma^{(\alpha)}}\right)^2}\right\rangle
\end{equation}

In the dilute limit, this quantity can be interpreted as the magnetic moment gained when adding a single vacancy. This is potentially useful for predicting phenomena such as defect-induced magnetism in irradiated SiC\cite{Zhou_2019}\cite{PhysRevLett.106.087205}.

\subsection{Vacancies and Order}

\begin{figure}[H]
    \centering
    \includegraphics[width=\linewidth]{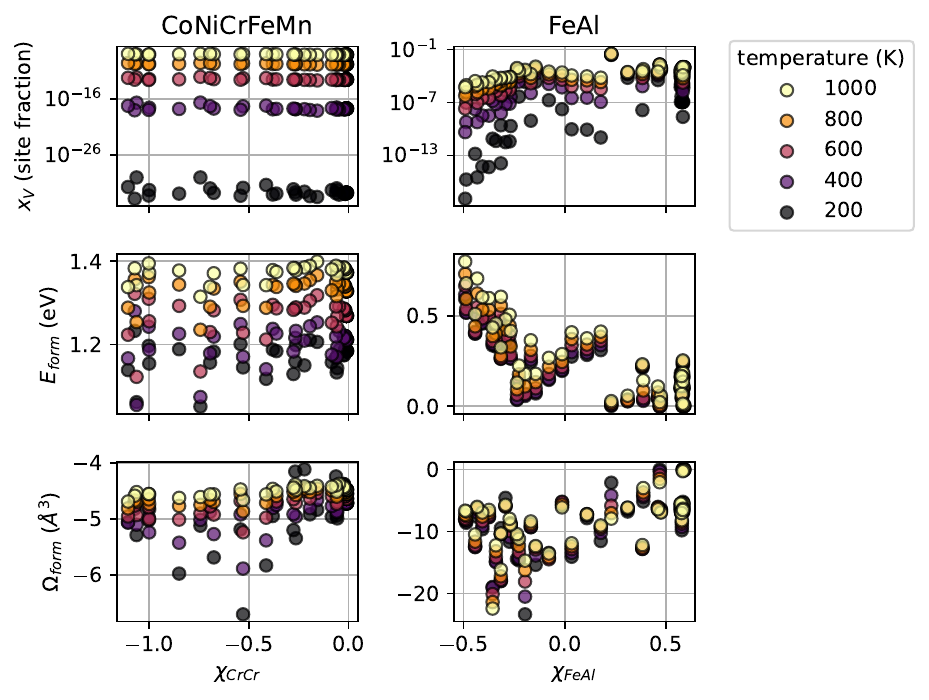}
    \caption{Vacancy concentration (top), formation energy (middle), and formation volume (bottom) as a function of an ordering parameter for various temperatures for the CoNiCrFeMn and FeAl (right) systems.}
    \label{fig:form-order}
\end{figure}

At each step that we compute $\chi_{\alpha\alpha'}$, we perform a vacancy insertion and substitutional swaps at each site. Then, we perform the associated thermodynamic calculations to compute vacancy concentration, formation energy, and formation volume for configurations with different ordering (see Figure \ref{fig:form-order}).

For the CoNiCrFeMn alloy, there is not a clear correlation between $\chi_{\text{Cr}\text{Cr}}$ and vacancy concentration, formation energy, nor formation volume at any tested temperature. However, for FeAl, vacancy concentration and formation volume increase with $\chi_{\text{Fe}\text{Al}}$, while formation energy decreases with $\chi_{\text{Fe}\text{Al}}$ (Figure \ref{fig:form-order}), with the trend between chemical ordering and vacancy concentration becoming flatter at high temperatures. In particular, vacancy concentration and formation energy reach extreme values (high concentration and small formation energy) for high $\chi_{\text{Fe}\text{Al}}$. Note that the high $\chi_{\text{Fe}\text{Al}}$ configurations, where Fe and Al are phase-separated, are very far out-of-equilibrium, leading to high-energy occupied states, and thus having similar energetics between vacant and occupied states. In comparison, the CoNiCrFeMn system is closer to equilibrium over the range of sampled ordering parameters, yielding a large energetic difference between vacant and occupied states.
\section{Limitations}

Our methodology does not account for binding between impurities/defects. While it should remain true that each site's occupation numbers follow the completely open ensemble and that the concentration of the impurity is the thermodynamic average of the impurity's occupation probability, we take a naive averaging procedure that does not account for correlations between sites, where $N$ is the number of lattice sites:

\begin{equation}
    \langle p_{(\sigma, \iota)}(\beta)\rangle = \frac{1}{N}\sum_\sigma p_{(\sigma, \iota)}(\beta)
\end{equation}

which can be problematic for many relevant solid solutions, such as those with a large number of divacancies, voids, and/or dislocations. Another example would be He in stainless steel, where He tends to precipitate \cite{doi:10.1080/01418618308245265}.

Even without a high tendency to bind, correlations between impurity occupation probabilities will be relevant outside of the dilute limit. However, at least for defects, the dilute limit is usually well-respected below the melting point \cite{SIEGEL1978117}\cite{KRAFTMAKHER199879}. In general, to include correlations, one must additionally include higher-order cluster terms:

\begin{equation}
    \begin{aligned}
        p_{(\sigma, \iota)}(\beta) &= \sum_{c\in C_m \text{ and } \sigma\in c}p_{(c,\iota)}(\beta)\\
        x_\iota(\beta) &= \frac{1}{N}\sum_\sigma p_{(\sigma, \iota)}(\beta)
    \end{aligned}
\end{equation}

where $C_m$ denotes the set of size $m$ clusters, and $p_{(c,\iota)}(\beta)$ denotes the joint occupation probability of $\iota$ within the cluster $c$.

Our numerical calculations in this work assumes the simplest case, i.e. $m = 1$. Derivations of $p_{(c, \iota)}(\beta)$ for larger values of $m$ are combinatorically complex, however. This combinatorial complexity is not only a problem for analytical derivations, but are also problematic for energetics computations, where the number of necessary energies induces combinatorial explosion. Furthermore, any symmetry argument to lower the number of computations is at least nontrivial, if not impossible, since the variance in atomic environments significantly lowers the symmetry of the problem. As such, this more general approach is saved for a future work.

Furthermore, sampling with a single well-equilibrated configuration might not be sufficient for small systems, since a small system might not well-sample different realizations of solid solution noise. This is problematic for an \textit{ab-initio} technique that introduces size constraints.

We also implicitly assume that energetics are independent of temperature. However, it has been predicted that local quasi-harmonicity and anharmonicity both play a crucial role in vacancy formation thermodynamics in Al and Cu at sufficiently high temperatures\cite{PhysRevX.4.011018}. Therefore, it might not be sufficient to only count microstates at $\SI{0}{K}$, but also vibrational states.

Lastly, the  CoNiCrFeMn interatomic potential used in this work was not fitted to vacancy energetics. As such, we expect a limited ability of quantitative prediction of our model using this potential. However, any energy-calculating method can be used to get the necessary data to predict atom/defect concentration. Using the MolSSI Driver Interface library\cite{BARNES2021107688}, one can couple an \textit{ab initio} method to LAMMPS for easy reuse of our prewritten input files and data analysis scripts. While prior work has shown that MC-MD methods driven by \textit{ab initio} energy calculations provide great insight into chemical ordering\cite{Widom2014}\cite{10.1063/5.0069417}, it is worth noting that using an \textit{ab initio} method is much more computationally expensive, and care must be taken to properly anneal a sufficiently large system.
\section{Conclusions}

In summary, we introduce a flexible single-site model for competitive element/defect occupation and use this model to predict element/defect concentration in any concentrated solid solution, including vacancy, small interstitial, and substitutional defects. This model can be modified to incorporate other thermodynamic forces, allowing for the prediction of impurity (element/defect) thermodynamics as a function of various experimental conditions. 

We then use this model to predict the vacancy concentration, formation energy, and formation volume of the CoNiCrFeMn and FeAl alloys at $\SI{0}{bar}$, and calculate all of these quantities as a function of short-range order, annealing from a random solution for the Cantor alloy and a segregated solution for FeAl. We show that, for the CoNiCrFeMn system over the $\SI{200}{K}$ to $\SI{1000}{K}$ range, there is no clear correlation between short-range order and vacancy thermodynamics. For the FeAl system over the same temperature range, however, vacancy formation energy decreases as the system moves further away from equilibrium chemical ordering, increasing vacancy concentration. We provide code and a corresponding tutorial to analyze and reproduce our results using LAMMPS, allowing for the quick evaluation of our model for a different use case.
\section{Data Availability}

All data, LAMMPS input files, post-processing scripts, and a brief tutorial for using the files and scripts are available on GitHub\cite{github_repo}\cite{harris2020array}\cite{Hunter:2007}.
\section{Acknowledgements}

Authors acknowledge support from the U.S. Department of Energy, Office of Basic Energy Sciences, Materials Science and Engineering Division under Award No. DE-SC0022980.

Additionally, this material is based on work supported by the National Science Foundation under Grant Nos. MRI\# 2024205, MRI\# 1725573, and CRI\# 2010270 for allotment of compute time on the Clemson University Palmetto Cluster.
\section{Disclaimer}

Any opinions, findings, and conclusions or recommendations expressed in this material are those of the author(s) and do not necessarily reflect the views of the National Science Foundation.
\appendix

\section{Appendix}

\subsection{Limits}\label{limits}

For substitutional impurities (atomic and vacancy), from Eq. \eqref{eq:local-prob}:

\begin{equation}
    \begin{aligned}
        \frac{1}{\sum_\alpha e^{\beta\mathcal{H}_\sigma^{(\alpha)}}} &= \frac{p_{(\sigma, \iota)}(\beta)}{1 - p_{(\sigma, \iota)}(\beta)}\\
        &= p_{(\sigma, \iota)}(\beta) + \mathcal{O}\left(\left[p_{(\sigma, \iota)}(\beta)\right]^2\right)
    \end{aligned}
\end{equation}

So, in the dilute limit, i.e. small $p_{(\sigma, \iota)}(\beta)$, the expressions for substitutional impurities (atomic and vacancy) are reduced to:

\begin{equation}\label{eq:small-sub}
    p_{(\sigma, \iota )}(\beta) \approx \frac{1}{\sum_\alpha e^{\beta\mathcal{H}_\sigma^{(\alpha)}}}
\end{equation}

and the expression for small impurities can be rewritten as:

\begin{equation}\label{eq:small-int}
    p_{(\sigma, \iota )}(\beta) = \frac{1}{\sum_\gamma e^{\beta \mathcal{H}_\sigma^{(\gamma)}}}
\end{equation}

where $\gamma\in\{\iota\}\cup\{(\iota, \iota')\;|\;\iota'\in\mathcal{I}\text{ and } \iota'\neq\iota\}$. Note that \eqref{eq:small-sub} and \eqref{eq:small-int} are only different in the sequence over which a summation is performed. Therefore, we can sum over $\mathcal{A}$ without loss of generality. In the dilute limit:

\begin{equation}
    \ln p_{(\sigma, \iota )}(\beta) = -\ln\left(\sum_\alpha e^{\beta\mathcal{H}_\sigma^{(\alpha)}}\right)
\end{equation}

Defining a random variable $\mathcal{H}_\sigma^{(\mathcal{A})}$ by a uniform distribution over $\{\mathcal{H}_\sigma^{(\alpha)}\;|\; \alpha\in\mathcal{A}\}$ yields:

\begin{equation}
    \begin{aligned}
        \ln p_{(\sigma, \iota )}(\beta)
        &=\ln\frac{1}{|\mathcal{A}|} - \ln \mathbb{E}\left[ e^{\beta\mathcal{H}_\sigma^{(\mathcal{A})}}\right]
    \end{aligned}
\end{equation}

Then $\ln \mathbb{E}\left[e^{\beta\mathcal{H}_\sigma^{(\mathcal{A})}}\right]$ is a cumulant generating function (CGF) of the random variable $\mathcal{H}_\sigma^{(\mathcal{A})}$ (assuming the random variable is temperature-independent), and can thus be expanded in powers of $\beta$ with cumulants $\kappa_\sigma^{(\ell)}$ as coefficients to some finite order $n$ \cite{brillinger_timeseries}:

\begin{equation}
    p_{(\sigma, \iota )}^{(n)}(\beta) = \frac{1}{|\mathcal{A}|}\exp\left( - \sum_{\ell=1}^n \frac{\kappa_\sigma^{(\ell)}}{\ell!}\beta^\ell\right)
\end{equation}

where $p_{(\sigma, \iota )}^{(n)}(\beta) \to p_{(\sigma, \iota )}(\beta)$. Then:

\begin{equation}
    x_\iota^{(n)}(\beta) = \frac{1}{|\mathcal{A}|} \left\langle\exp\left( - \sum_{\ell=1}^n \frac{\kappa_\sigma^{(\ell)}}{\ell!}\beta^\ell\right)\right\rangle
\end{equation}

where $x_\iota^{(n)}(\beta)\to x_\iota(\beta)$. Define the $n$-dimensional random variable $\mathbf{k}_\sigma$ as:

\begin{equation}
    \mathbf{k}_\sigma = -\left(\kappa_\sigma^{(1)}, \frac{\kappa_\sigma^{(2)}}{2}, \cdots, \frac{\kappa_\sigma^{(n)}}{n!}\right)
\end{equation}

and the variable $\mathbf{b}(\beta)$ as:

\begin{equation}
    \mathbf{b} = \left(\beta, \beta^2, \cdots, \beta^n\right)
\end{equation}

Therefore:

\begin{equation}
    \ln x_\iota^{(n)}(\beta) = \ln\frac{1}{|\mathcal{A}|} + \ln\left\langle e^{\mathbf{b}(\beta)\cdot \mathbf{k}_\sigma}\right\rangle
\end{equation}

where $\ln\left\langle e^{\mathbf{b}(\beta)\cdot\mathbf{k}_\sigma}\right\rangle$ is another CGF. Expanding this CGF about $\mathbf{b}(\beta) = \mathbf{0}$ gives:

\begin{equation}
    \begin{aligned}
        \ln x_\iota^{(n)}(\beta) &= \ln\frac{1}{|\mathcal{A}|} + \left\langle\mathbf{k}_\sigma\right\rangle\cdot \mathbf{b}(\beta)\\
        &+ \frac{1}{2}\mathbf{b}(\beta)\cdot\mathbf{K}\mathbf{b}(\beta) + \cdots
    \end{aligned}
\end{equation}

where $\mathbf{K}$ is the covariance matrix with elements $K_{\ell\ell'}$:

\begin{equation}
    K_{\ell\ell'} = \frac{1}{\ell!\ell'!}\left\langle \left(\kappa_\sigma^{(\ell)} - \left\langle\kappa_\sigma^{(\ell)}\right\rangle\right)\left(\kappa_\sigma^{(\ell')} - \left\langle \kappa_\sigma^{(\ell')}\right\rangle\right)\right\rangle
\end{equation}

Therefore:

\begin{equation}
    x_\iota^{(n)}(\beta) = \frac{1}{|\mathcal{A}|}e^{-\beta\left\langle\kappa_\sigma^{(1)}\right\rangle -\frac{1}{2}\beta^2\left(\left\langle\kappa_\sigma^{(2)}\right\rangle - K_{11}\right) + \mathcal{O}\left(\beta^3\right)}
\end{equation}

So, in the high-temperature limit, we recover an Arrhenius expression with activation energy $\left\langle\kappa_\sigma^{(1)}\right\rangle =\left\langle\frac{1}{|\mathcal{A}|}\sum_\alpha\mathcal{H}_\sigma^{(\alpha)}\right\rangle$.

\subsection{Supercell Method}\label{supercell}

The vacancy and anti-site defect formation energies on both sublattices can be calculated from the supercell method. We calculate these by creating a perfect $5\times 5\times 5$ B2 supercell, minimizing to $\SI{0}{bar}$, and evaluating the energy $E_0$.

We then swap an Al atom with an Fe atom and minimize, calculating $E_{\text{Al}\to\text{Fe}}$, and similarly calculate $E_{\text{Fe}\to\text{Al}}$. We then return to the pristine sample, and remove an arbitrary Al atom to create a vacancy, minimizing and evaluating the energy $E_\text{Al}^*$. We similarly calculate $E_\text{Fe}^*$.

\begin{table}[H]
    \centering
    \begin{tabular}{|c|c|}
            \hline
         $E^\circ$ & $ \SI{-1030.8746}{eV}$  \\
         \hline
         $E_{\text{Al}\to\text{Fe}}$ & $\SI{-1031.4861}{eV}$\\ \hline
        $E_{\text{Fe}\to\text{Al}}$ & $\SI{-1029.0932}{eV}$ \\ \hline
         $E_{\text{Al}}^*$ & $\SI{-1025.837}{eV}$ \\ \hline
        $E_{\text{Fe}}^*$ & $\SI{-1024.861}{eV}$ \\
        \hline
    \end{tabular}
    \caption{Swapping energies from supercell method}
    \label{tab:swaps}
\end{table}

The vacancy formation energies on both sublattices are then:

\begin{equation}
    \begin{aligned}
        \Delta E_{\text{Al sublattice}}^\text{v} &= E_\text{Al}^* - E^\circ + \mu_\text{Al} = \SI{1.54}{eV} \\
        \Delta E_{\text{Fe sublattice}}^\text{v} &= E_\text{Fe}^* - E^\circ + \mu_\text{Fe} = \SI{1.40}{eV}
    \end{aligned}
\end{equation}

and the anti-site formation energies are:

\begin{equation}
    \begin{aligned}
        \Delta E_{\text{Al sublattice}}^\text{Fe} &= E_{\text{Al}\to\text{Fe}} - E^\circ + \mu_\text{Al} - \mu_\text{Fe}\\
        &=\SI{0.51}{eV} \\
        \Delta E_{\text{Fe sublattice}}^\text{Al} &= E_{\text{Fe}\to\text{Al}} - E^\circ + \mu_\text{Fe} - \mu_\text{Al}\\
        &=\SI{0.66}{eV}
    \end{aligned}
\end{equation}

\bibliography{bibfile.bib}

\end{document}